\documentclass{article}

\usepackage{arxiv}

\usepackage[T1]{fontenc}    

\usepackage{url}            
\usepackage{booktabs}       
\usepackage{amsfonts}       
\usepackage{nicefrac}       
\usepackage{microtype}      
\usepackage{graphicx}
\usepackage{natbib}
\usepackage{doi}
\usepackage{array} 
\usepackage{parskip}
\usepackage{booktabs}
\usepackage{color}
\hypersetup{
  colorlinks   = true, 
  urlcolor     = blue, 
  linkcolor    = blue, 
  citecolor   = blue 
}

\title{SARD: A Human-AI Collaborative Story Generation}

\date{} 					

\author{\hspace{1mm}Ahmed Y. Radwan \\
	Department of Computer Science\\
    ASAS AI \\
	King Abdulaziz University\\
	\texttt{aragabradwan@stu.kau.edu.sa} \\
    \And
    \hspace{1mm}Khaled M. Alasmari \\
	Department of Computer Science\\
    ASAS AI \\
	King Abdulaziz University\\
	\texttt{kabdualasmari@stu.kau.edu.sa} \\
    \And
    \hspace{1mm}Omar A. Abdulbagi \\
	Department of Computer Science\\
    ASAS AI \\
	King Abdulaziz University\\
	\texttt{ofaridabdulbagi@stu.kau.edu.sa} \\
    \And
    \hspace{1mm}Emad A. Alghamdi\\
	Center for Research Excellence in AI and Data Science\\
    ASAS AI\\
	King Abdulaziz University\\
	\texttt{eaalghamdi@kau.edu.sa} \\
}

\date{}

\renewcommand{\shorttitle}{\textit{Sard}}

\begin{document}
\maketitle

\begin{abstract}
Generative artificial intelligence (GenAI) has ushered in a new era for storytellers, providing a powerful tool to ignite creativity and explore uncharted narrative territories. As technology continues to advance, the synergy between human creativity and AI-generated content holds the potential to redefine the landscape of storytelling. In this work, we propose SARD, a drag-and-drop visual interface for generating a multi-chapter story using large language models. Our evaluation of the usability of SARD and its creativity support shows that while node-based visualization of the narrative may help writers build a mental model, it exerts unnecessary mental overhead to the writer and becomes a source of distraction as the story becomes more elaborated. We also found that AI generates stories that are less lexically diverse, irrespective of the complexity of the story. We identified some patterns and limitations of our tool that can guide the development of future human-AI co-writing tools.
\end{abstract}


\section{Introduction}

Narrative expression is a fundamental element of the human experience, manifested across various mediums such as written text, oral traditions, cave paintings, and beyond. Generative artificial intelligence has ushered in a new era for storytellers, providing a powerful tool to ignite creativity and explore uncharted narrative territories \citep{kybartas2016survey, lee2024design}. Generative AI, a subset of artificial intelligence, has revolutionized the way stories are crafted, providing a new avenue for creativity and imagination. While extensive work on automated story generation has been conducted in the past 40 years \citep{kybartas2016survey}, the rise of large language models (LLMs) is opening new frontiers for human-AI collaboration. However, integrating such capabilities with human cognitive faculties and creative processes remains challenging. 

Generative AI, often associated with deep learning techniques, enables machines to produce content that mimics human-like creativity. Unlike traditional programming, which relies on explicit instructions, generative AI learns patterns and styles from vast datasets to create original and diverse content. This technology has found a unique niche in the world of storytelling, where it is employed to generate narratives, characters, and even entire stories and poems \citep{chakrabarty2022help}. Generative AI for story creation involves training models on massive datasets comprised of diverse literary works, genres, and writing styles. These models, such as OpenAI's GPT-3 and GPT-4, learn the nuances of language, syntax, and semantics, enabling them to generate coherent and contextually relevant text. Once trained, these models can be prompted with a starting point or theme, and they autonomously produce imaginative and contextually appropriate stories.

One of the significant advantages of generative AI in story creation is its ability to break through creative barriers and expedite the writing process \citep{lee2024design}. Exposing the model to a myriad of writing styles and genres can produce content combining elements from various sources, resulting in novel and unique narratives. This not only aids writers in overcoming creative blocks but also fosters a collaborative relationship between human creativity and machine-generated content. Generative AI should be viewed as a tool to augment human creativity rather than replace it. Writers can leverage these AI models to explore new ideas, overcome writer's block, or even collaborate with the AI in co-authoring projects. The human touch remains essential in refining and contextualizing the generated content, ensuring that it aligns with the intended emotional tone, narrative structure, and thematic elements.


\section{Related Work}
The use of computer programs to generate a story or parts of a story has been an interest for computer science researchers since the field's inception \citep{kybartas2016survey}. Initial efforts in this field utilized traditional AI algorithms, including symbolic and logical planning as well as graph traversal, to craft narratives \citep{klein1973automatic, dehn1981story}. These narratives often incorporated a degree of user control, allowing users to define initial goals and conditions. In recent endeavors,  researchers have explored the potential of using LLMs to produce complete narratives autonomously \citep{fan2018hierarchical}, whereas others have emphasized the significance of developing AI systems that give precedence to human participation in the process of crafting stories \citep{swanson2021story}. 

\subsection{Generative AI for Story Creation}
Recent advancements in generative AI, such as GPT-4 and DALL-E2, have unlocked new possibilities for automated story generation. Numerous prototypes of AI-powered story-authoring systems have emerged. \cite{yuan2022wordcraft} developed Wordcraft, a web application for story writing with an LLM. Wordcraft consists of a traditional text editor and a set of controls that prompt an LLM to perform various writing tasks. In \cite{yang2022ai}, writers interacted with AI in a "turn=taking" style to co=write short science-fiction stories. The writers discovered that the language model occasionally produces texts of subpar quality, containing words that may be challenging to comprehend. On the contrary, there are instances where it generates high-quality inspirations that propel the plot beyond what humans might anticipate. \cite{calderwood2020novelists} explored how novelists use AI to generate stories finding that novelists find pleasure in utilizing the language model as a constraining tool to push the boundaries of their writing or as an opponent that aided them in realigning and improving their purpose. \cite{osone2021buncho} investigated AI-assisted storytelling for Japanese novelists and found that while novice writers benefited from model suggestions, experts found these less useful and underwhelming. 

Concerns about LLM's creativity have been raised. \cite{padmakumar2023does} found that LLMs aligned to human feedback, e.g., InstructGPT, generate less diverse content and increase the similarity between the writings of different writers. Addressing the lack of long-semantic coherence and relevance of outputs generated by LLMs, \cite{yang2022re3} proposed the Recursive Reprompting and Revision framework (Re3) which recursively prompt LLMs to plan, draft, rewrite, and edit. A qualitative evaluation by human evaluators found stories generated using this framework to have more coherent plots and relevance to the user install premise. \cite{mirowski2023co} developed Drematron by applying language models hierarchically via prompt chaining. Drematron can help theatre and film professionals generate coherent scripts and screenplays along with titles, characters, story beats, location descriptions, and dialogues. 

\subsection{AI-Authoring Tools}
There has been some recent work on integrating generative AI in writing assistant systems and how to best support storytellers in their writing process. These tools vary in their level of technical complexity, degrees of automation, user interactions, and level of support they provide. A common design paradigm for writing authoring tools takes the form of dialogue, where a user and the language model take turns to append content to the end of the story \citep{calderwood2020novelists, yang2022ai, yuan2022wordcraft}. For example, \cite{clark2018creative} performed a case study on collaborative slogans and short story writing, observing that users preferred an interface design where they had a higher level of control over the interactions.
\subsection{AI Content Quality and Novelty}
Assessing the quality of content generated by AI models remains a difficult challenge, as tasks of quantifying the intricacies of narrative structure, creativity, coherence, and emotional resonance in AI-generated stories are complex. Numerous studies have approached this evaluation by assessing their generated stories with various evaluation methods. One of the most efficient methods is computer-based evaluation, where the stories generated are automatically assessed for quality, relevance, and coherence using tools and algorithms. Studies, such as \cite{kong2021stylized} and \cite{khan2023learning}, used pre-trained models to evaluate the fluency and diversity in AI-generated content.

\section{System Overview}
\begin{figure}[ht]
    \centering
    \includegraphics[height=0.55\textwidth, width=0.75\textwidth]{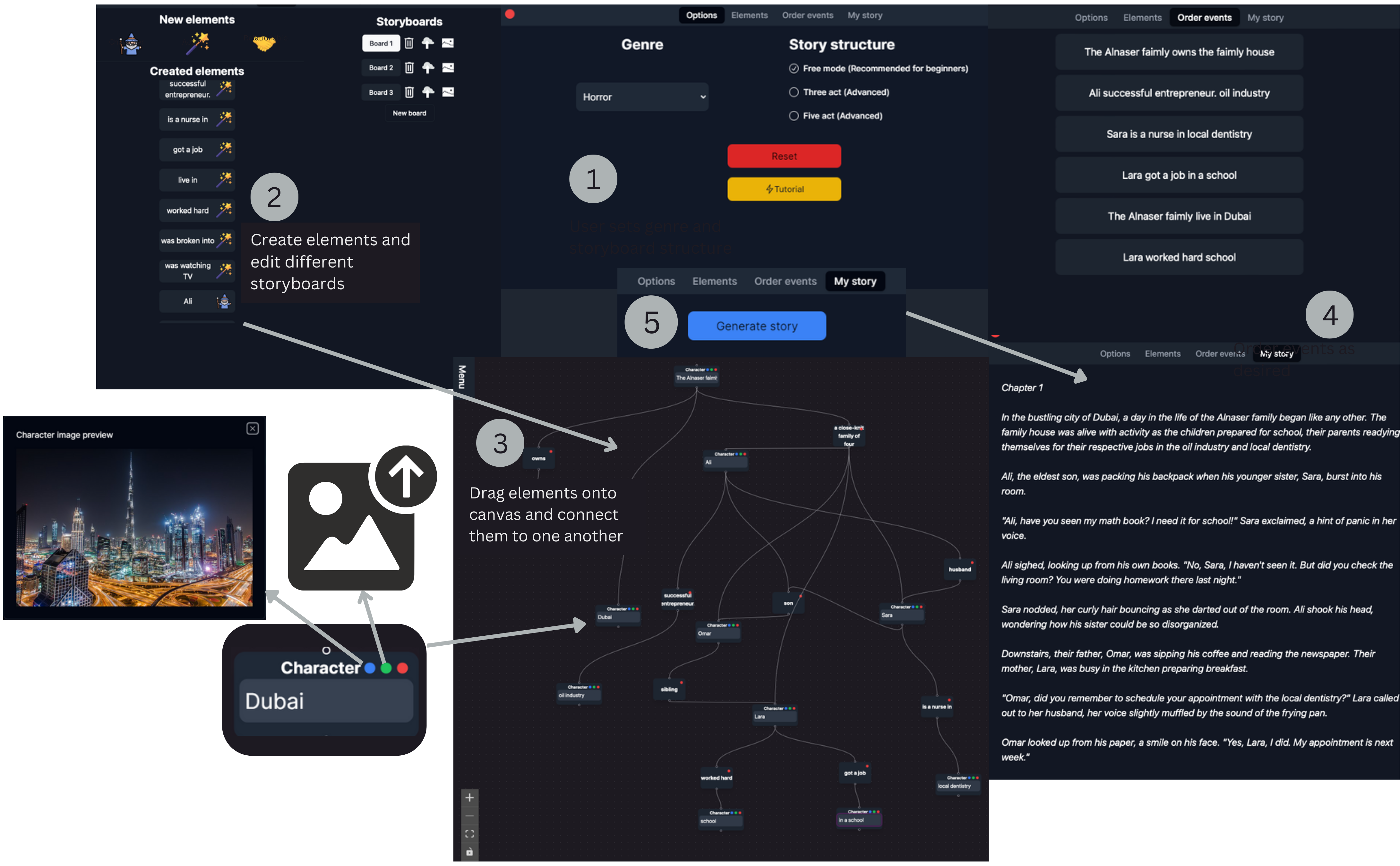}
    \caption{SARD: Novel generation through interactive storyboard. The user starts with (1) where the user sets the story structure and genre. (2) is where the user can create and add elements, either character, action, or relationship, to their storyboard. (2) can also be used to navigate through different storyboards and set the setting with an image. When creating an element the user drags it into the canvas (3) where it is translated into a node. The user can click the green button to represent character nodes with images, and then preview them with the blue button. This node can be connected to other elements creating relationships. After the user orders events (4), they can generate their storyboards into a story (5) which is displayed in the story tab. 
    }
    \label{fig:sard}
\end{figure}

SARD is a multi-chapter story generation facilitated by generative AI (see \ref{fig:sard}). It is a storyboard-based authoring tool that enables users to construct narratives through a drag-and-drop interface. SARD editor is built with ReactJS and shadcn/ui to help users to generate their own stories via a simple drag-and-drop interface. SARD is connected to generative AI models through a REST API and a WebSocket connection to the backend. 

Users initiate the creation of a story by accessing the menu and selecting their preferred story genre (thrillers, science fiction, ..etc ) and structure (free, three-act story, and five-act story). Once initial story parameters are set, the users can add nodes, provided in the story elements tab, that symbolizes different narrative components such as characters, actions, and relationships. These nodes can be interconnected with one another to form a complex narrative structure. Another key feature of SARD is the generation of descriptive content for characters or scenery in the story based on images provided by the user. After finishing the storyboard, the users can order events to enhance the narrative flow. The finalized user-generated storyboard is converted programmaticly to prompts which are then sent to the GPT-4 model to craft a coherent and contextually rich story. 
\subsection{Prompt Designing}
The designed several prompts for the different functionalities in SARD. Table \ref{table:promots} shows the prompts and their functions. We experimented with different prompts for each functionality and only included the ones generated the best results. We designed our prompts to be as generic as possible. For example, our prompt for ensuring coherence between all chapters is "\texttt{Summarize the following chapter in a short and concise way. Make sure you include all the important events in your summary as the next chapters will depend on it: [chapter to summarize]}." We observed that this prompt helped the language model to grasp what happened in the preceding and following events. We decided to hide the prompts from the user to make them focus more on their story-writing process and less on optimizing prompts. Our decision was motivated by a previous study in which non-AI experts found it difficult to craft suitable prompts to generate intended behaviors \citep{zamfirescu2023johnny}.

\begin{table}
 \begin{tabular}{|>{\centering\arraybackslash}p{0.2\linewidth}|>{\itshape\raggedright\arraybackslash}p{0.7\linewidth}|}\hline Functionality & Prompt\\

\hline 
          Describing a scenery & \texttt {This image represents a place where events happened. Describe the place in detail. If the image has characters in it, do not describe them and ignore them. Your main focus is to describe the place and its surroundings in detail}.\\ \hline 
         
         Describing the visual appearance of a character & \texttt{Describe the character's appearance in detail for the attached image. The character name is \colorbox{cyan}{[name]}, refrain from using pronouns, please use the character name instead. Make sure you start describing the character immediately. Do not use words like 'Certainly', or 'Okay'. If you do not receive an image, respond with nothing}.\\ \hline 

Generating a story & \texttt{Write chapter \colorbox{cyan}{[chapter number]} with dialogues using the following \underline{characters' details:}\newline
[characters names and details]\newline
and the relationships between them are \colorbox{cyan}{[list of relations]}.\newline
now map it to the information you have in the following events:\newline
\colorbox{cyan}{[list of events]}
\colorbox{cyan}{[previous chapters summary]}\newline
<\colorbox{cyan}{[description for the place where the events happened]}>"""\newline
\underline{Output Length:} """<3000 words>"""\newline
\underline{Structure of Writing:} """<You are writing a chapter, follow the rules to write an amazing chapter"""\newline
Take your time with the writing, perfecting this chapter}.\\ \hline

Summarizing a chapter & \texttt{Summarize the following the chapter shortly and concisely. Make sure you include all the important events in your summary as the next chapters will depend on it:\newline
\colorbox{cyan}{[chapter to summarize]}}\\ \hline 
    \end{tabular}
    \caption{The prompts used to instruct GPT-4}
    \label{table:promots}
\end{table}

\subsection{Canvas}
The canvas resembles the storyboard where story characters and different elements will be visualized as nodes. Specific nodes can be linked together through edges to form an event in the current storyboard. In addition, the optional initialization of metadata by providing character nodes with images. The simplistic interface is aimed to be user-friendly with a mini-map feature and a navigational guide to freely transport through the canvas. These functionalities streamline the user's creative workflow and provide a clear picture of how the story is developed progressively.

\subsection{Setting Genre and Structure}
The options tab offers two main components to writing a novel, the first one is the genre list and the second is the story structure. The genre's list provides our model to be more strict, as in the novels each genre has a way of writing and leads to a different flow of events and words. For the story structure, there are three options, free mode which lets the user have no rules in writing the story, as he may have at least one board to generate a story. There is no maximum that will make the story more creative but with the disadvantage of not following a famous structure like the others. The three-act structure consists of three main parts, an introduction, a climax, and a resolution, to represent each one in a storyboard, as it will control the flow of the story, unlike the free mode. And, for the last structure is the five-act structure, which follows \textit{Exposition, Rising, Climax, Falling Action}, and \textit{Resolution}. The main difference from the three-act structure it will be longer which will be more engaging in longer stories.

\subsection{Adding Characters, Events and Actions}
The story is built with three types of nodes: \textit{character, action}, and \textit{relationship}. The user can create one or more storyboards and drop new elements or reuse any existing story elements. The character node allows the user to add new characters as well as images to describe the character or scenes. For the action node, it offers multiple default actions to choose from, as well as a custom option that allows the user to be more specific about what he wants. Finally, the relation node is intended to bring dimension to the characters. These components are used to build the novel's events, which each include at least one character on both sides and are connected by an action node or relation. For example, the event 'Ahmad humiliated John and Ben' will be divided into four nodes combined from three characters, and a single action. 

\subsection{Ordering of Events}
The events sequence is of importance as it dictates the flow and coherence of the story. Based on user feedback and testing through development, we discovered that chaining events with edges can get messy and confusing very quickly. To overcome this, we added a tab for ordering events in the interface.

\section{Methodology}
\subsection{Study 1: Evaluating The usability and collaborative aspects of SARD}
 The primary objective of Study 1 was to assess the usability of SARD and its creativity support. To this end, the System Usability Scale (SUS) \citep{brooke1996sus} and a modified version of the Mixed-Initiative Creativity Support Index (MICSI) \citep{lawton2023drawing} were used. The SUS scale includes 10 items (five-point Likert scale questionnaire), developed specifically to provide a reliable metric for system usability. The final score of SUS ranges from 0 to 100 and a score of 70 or above is considered to be a good indicator of system usability \citep{bangor2009determining}. MICSI, an 18-item scale designed for evaluating collaborative creativity between humans and machines, encompasses five sub-scales related to Creativity Support: Enjoyment, Exploration, Expressiveness, Immersion, and Results-Worth-Effort. Each of these sub-scales involves a pair of seven-point Likert-scale questions, with the score calculated based on the mean response value for each question pair. Additionally, MICSI includes sub-scales dedicated to assessing Human-Machine Collaboration, each measured through a distinct seven-point Likert-scale question. These sub-scales address aspects like Communication, Alignment, Agency, and Partnership. A score of 5 or higher on these MICSI sub-scales indicates a positive user experience. We also allowed users to give any feedback on their use of SARD. 
 
\subsubsection{Participants}
 We recruited five university students (4 females and 2 males) to participate in the study. All the participants were graduate students in English and creative writing. The participants were aged between 22 and 25 (\textit{M}= 22.5, \textit{SD} = 0.42).
 
\subsubsection{Task and Procedure}
After obtaining participants' consent, they were given a walk-through tutorial on how to navigate the interface and use the system. Then, they were asked to generate a story of their choice using all the utilities of SARD. They were given a week to complete a story. 

\subsubsection{Results} 
\begin{figure}[ht]
    \centering
    \includegraphics[height=0.35\textwidth, width=1.0\textwidth]{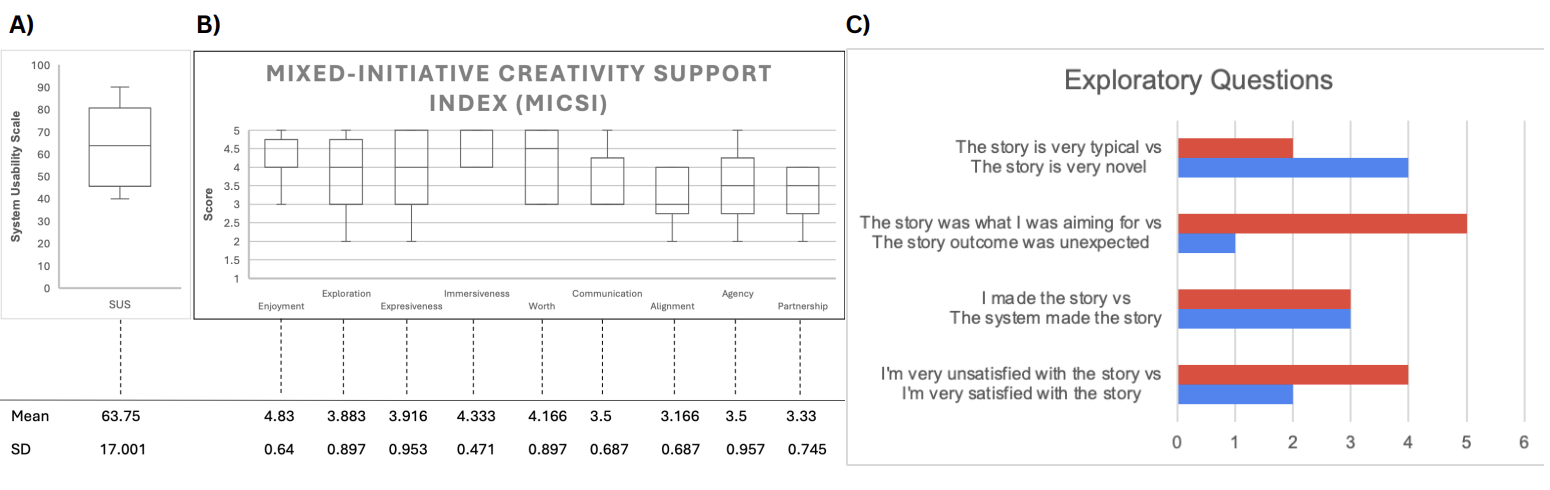}
    \caption{Results from Study 1: (a) SUS Scores, (b) MICSI Sub-Scale Scores, (c) Exploratory Question Responses.}
    \label{fig:results}
\end{figure}

The results of scales are presented in \ref{fig:results}. The average score for SARD's usability was 63.75, indicating that participants had moderately positive attitudes toward the usability of SARD. This could be due to the novelty effect. Two participants indicated they initially had difficulty connecting the nodes in the canvas. 

\indent \textbf{W1} \textit{"I think how to connect the nodes was a bit challenging at first. It's hard to link nodes with each other but after some attempts, I was able to build my story "}

We also observed that some participants did not go through the tutorial we provided in the interface, and immediately started adding characters to the canvas. 

The scores of the MICSI sub-scales were slightly below five, as shown in \ref{fig:results} B, but they all expressed their satisfaction with their generated stories as they reflected in the exploratory questions as well as in their feedback in Study 2. 

\subsection{Study 2: Evaluating Story Quality}
 Understanding how writers assess systems output is crucial because it could influence their interaction and engagement with the writing assistant during the story generation process. One key characteristic of high-quality output is its textual coherence, which underlines the need for grammatically and contextually coherent output \cite{gero2022sparks}. Another important facet is textual diversity which underscores the importance of offering varied system outputs to foster creativity in writing \cite{singh2023hide, gero2022sparks}. In Study 2, we investigated the quality and content diversity of the generated stories via quantitative and qualitative measures. 
\subsubsection{Measures}
We assessed the lexical diversity of the generated stories using token-type ratios (TTR). We also conducted a semi-structured interview with the participants to solicit their opinions about the stories they generated using SARD. 

\subsection{Results}
All participants selected a three-act story structure and developed three storyboards for each act. Our analysis of the lexical diversity of generated stories showed that the participants' stories had similar TTR (\textit{M} = 0.47, \textit{SD} = 0.09). 

Regarding the participants' judgment of their own stories, all participants expressed their positive attitudes toward the story they generated using SARD. Specifically, they mentioned how the tool helped them to elaborate on their initial plot and extend their vocabulary. \newline

\indent \textbf{W2} \textit{"I like it, it creates a very advanced level of the story. It uses advanced vocabulary."}

\indent \textbf{W5} \textit{"Honestly, I was not expecting that. I had a simple story in my mind but this tool took it to another level."}

\section{Discussion}
The results of our questionnaire and interviews with the participants have revealed some common patterns and themes in users' interactions with SARD, which can provide some insights for research in the realms of Natural Language Processing and Human-Computer Interaction.
\subsection{Identifying Patterns in User Interactions}
All writers found creating nodes to create relationships between the characters and events to help create a mental model of the story. However, a writer mentioned that she began to lose sight of her mental model of the story as she elaborated more on her story and added too many characters and events to her story canvas. While SARD summarizes and generates a plot for each chapter, it does not show it to the users. Showing a concise summary for each chapter might be helpful for users to stay oriented in their story generation process as it follows. We also observed that our participants developed their stores' characters and modified their plots repeatedly over a week. This may indicate that while our node-based system gives more flexibility and elaborateness to the story development process, it nonetheless exerts more cognitive load compared to turn-taking authoring systems.

\subsection{Sense of agency, control, and ownership}
The user's different perceptions of AI in the co-writing process seem to impact how they interact with the tool. While some users rely heavily on AI to steer the story generation, others want more control of their story. Specifically, writers felt AI does not reflect and express what they had in their mind, despite their efforts in planning the plot of the story and how the events should flow. Some participants felt that hiding the prompts used for generating stories is limiting their ability to control their stories.


\subsection{System Limitations Experienced by Participants}
The participants identified some limitations in SARD. First, some participants hoped that AI could help them in brainstorming ideas and initiating some suggested plots to expiate the writing process. A participant mentioned that he used ChatGPT to brainstorm some ideas before using SARD. Second, some participants expressed disappointment or resignation when the system's output did not align with their expectations, and that adjusting nodes and reordering the event sequence did not help generate more diverse content. This echoes recent concerns on the inability of current LLMs to generate creative content on their own \citep{chakrabarty2023art, franceschelli2023creativity}.

\section{Conclusion and Future Work}
AI-driven storytelling is an evolving landscape. In this study, we reported the findings of two studies on the usability and quality of AI-powered story generation using a custom-built interface. Our results demonstrated that the potential of human-AI co-creative systems, such as SARD, in helping novice writers generate novel stories. However, future systems need to take user's workload and their expectations of such systems into consideration. In our future work, we aim to give users more control over their story generation by allowing them to directly prompt and co-create with AI as a companion. We also aim to leverage LLMs for generating graph nodes from natural text describing story plots. 

\bibliographystyle{unsrtnat}
\bibliography{references} 

\end{document}